\documentclass [12pt]{article}
\usepackage {graphicx}

\begin{document}

\title{\bf Small viscosity method and criteria for shock wave existence in
relativistic magnetic hydrodynamics}

\author{V.I.Zhdanov, M.S.Borshch\\
\small Kyiv National Taras Shevchenko University\\
 }

\date{}

\maketitle

\begin{center}
\textbf{Abstract.}
\end{center}

 {\bf We obtain criteria for shock wave (SW)
existence in relativistic magnetic hydrodynamics with with no
suppositions about convexity of the equation of state. Method of
derivation involves consideration of a continuous SW profile in
presence of Landau-Lifshitz relativistic viscosity tensor with
both non-zero viscosity coefficients $\eta $ and $\zeta $. We
point out that supposition of viscous profile existence with only
one nonzero coefficient ($\eta $=0) appears to be too restrictive
leading to losses of some physical solutions. }

PASC classification codes

47.75.+f Relativistic fluid dynamics

52.35.Tc Shock waves

Keywords: relativistic magnetohydrodynamics, shocks, general
equation of state

\section{Introduction}

Relativistic shock waves (SW) arise in such powerful astrophysical
phenomena as supernova explosions and gamma-ray bursts.
Theoretical analysis of these processes involves criteria of
existence and stability of discontinuous solutions that describe
SW in superdense matter. Consideration of these criteria is
complicated in case of general equation of state (EOS) (cf., e.g.,
the classical results \cite{Rozhe} and the relativistic
hydrodynamics \cite{BGZh,BGKZh}).

Note that in case of a normal fluid we deal with convex EOS (this
means the convexity of Poisson adiabats); therefore the only
condition is needed to study discontinuous solutions of
hydrodynamical equations: the well known entropy growth criterion
(see \cite{Rozhe, LLFl} for classical hydrodynamics and
\cite{Lich, Sibg} in relativistic MHD). However, in case of a
general EOS the convexity condition may be violated, and neither
customary entropy criterion,  nor the evolutionarity criterion
\cite{LLFl} are not sufficient to single out physical solutions in
a correct way. Moreover, the rarefaction shocks and the
compression simple waves, as well as complicated configurations of
shocks and simple waves moving in the same direction are possible.
This situation is well-known in classical hydrodynamics, it was
first studied by H.Bethe  \cite{Rozhe}. In the relativistic theory
such anomalous equations of state arise, e.g., when dealing with a
super-dense matter in the neighborhood of phase transitions (see,
e.g., \cite{BGZh,BGKZh}).

One of the most effective methods to study the SW existence in
case of the general EOS is investigation of the SW viscous
profile. According to this method the generalized (discontinuous)
solution is treated as a small viscosity limit of corresponding
continuous solutions. The shock transition is admissible, if
corresponding continuous solution (viscous profile) exists for any
nonzero viscosity. In case of normal fluid (in the sense of Bethe
and Weyl \cite{Rozhe}) the results of this method are the same as
that of the evolutionarity criterion. In the relativistic
hydrodynamics the conditions for viscous profile existence in case
of the general EOS have been derived and studied \cite{BGKZh,BGZh}
by using the Landau-Liftshits viscosity term in relativistic
energy-momentum tensor \cite{LLFl}. This term involves two
viscosity coefficients $\xi $ and $\eta $.

In relativistic magnetic hydrodynamics (MHD)  \cite{Lich,Sibg}
investigation of the SW viscous profile becomes more complicated.
Therefore this problem has been first considered
\cite{ZhTytPLA,ZhTytZhTJE} in a restricted version with one of the
viscosity coefficients put equal to zero ($\eta =0 $) under
supposition that only one non-zero viscosity is sufficient to
obtain the continuous SW profile. This was a technical supposition
and it is not evident. At least, the results of
\cite{ZhTytPLA,ZhTytZhTJE} for $\eta =0 $ cannot be considered as
necessary conditions.

In the present paper we extend the results of
\cite{ZhTytPLA,ZhTytZhTJE} to the case of arbitrary ratio of
positive viscosity coefficients and prove conditions for existence
of the SW viscous profile under less restrictive requirements. We
consider stationary viscous flows of relativistic fluid with
infinite conductivity. These solutions describe the MHD shock
structure, the existence of SW being considered by means of
corresponding continuous solutions with non-zero viscosity. Our
treatment shows that we may relax the conditions of
\cite{ZhTytPLA,ZhTytZhTJE} to have a necessary and sufficient
criteria.

\section{Basic equations}

The equations of motion of ideal relativistic fluid with infinite
conductivity in magnetic field follow from the conservation laws
involving the energy-momentum tensor \cite{Lich,Sibg}

%(1)
\begin{equation}
\label{eq1} T^{\mu \nu}=(p^{*}+\varepsilon^{*})u^{\mu}
u^{\nu}-p^{*}g^{\mu \nu}-\frac{\mu}{4\pi}h^{\mu}h^{\nu} ,
\end{equation}

$u^{\mu}$ is the four velocity (Greek indexes run from 0 to 3),
the flat space-time metric $g^{\mu \nu} =g_{\mu \nu}
=diag(1,-1,-1,-1)$ is used for raising and lowering the indexes,
$h^{\mu}  = - \frac{1}{2}e^{\mu \alpha \beta \gamma} F_{\alpha
\beta}  u_{\gamma}  $ is the magnetic field,$ e^{\alpha \beta
\gamma \delta} $ is the absolutely anti-symmetric symbol, $F_{\mu
\nu} $ is the tensor of electromagnetic field, $p^{*} = p +
\frac{\mu}{8\pi }{\left| {h} \right|}^{2}$, $\varepsilon ^{ *} =
\varepsilon + \frac{\mu}{8\pi }{\left| {h} \right|}^{2}$, ${\left|
{h} \right|}^{2} = - h^{\alpha} h_{\alpha}  > 0$, $\mu $ is the
magnetic permeability that is supposed to be constant; $p$ is the
pressure and $\varepsilon $ is the energy density (in the rest
frame). We suppose EOS $p=p(\varepsilon ,n)$ to be a sufficiently
smooth function.

Following the small viscosity method
\cite{Rozhe,BGKZh,ZhTytPLA,ZhTytZhTJE} in order to study the SW
structure we introduce dissipation effects that smear out
discontinuities. Similarly to
\cite{BGZh,BGKZh,ZhTytPLA,ZhTytZhTJE}, in case of relativistic
problem we use the Landau-Liftshits viscosity tensor \cite{LLFl}

\[
\tau _{\mu \nu}  = \eta (u_{\mu ,\nu}  + u_{\nu ,\mu}  - u_{\mu}  u^{\alpha
}u_{\nu ,\alpha}  - u_{\nu}  u^{\alpha} u_{\mu ,\alpha}  ) + (\xi - 2\eta /
3)u^{\alpha} _{,\alpha}  (g_{\mu \nu}  - u_{\mu}  u_{\nu}  ),
\]

\noindent
the commas stand for derivatives.

Now the fluid motion is constrained by equations of energy-momentum
conservation

\begin{equation}
\label{eq2}
\partial _{\mu}  (T^{\mu \nu}  + \tau ^{\mu \nu} ) = 0,
\end{equation}

\noindent baryon charge conservation

\begin{equation}
\label{eq3}
\partial _{\mu}  (nu^{\mu} ) = 0,
\end{equation}

\noindent and one more equation follows from the Maxwell's
equations \cite{Lich,Sibg}

\begin{equation}
\label{eq4}
\partial _{\mu}  (u^{\mu} h^{\nu}  - u^{\nu} h^{\mu} ) = 0  .
\end{equation}

The discontinuous solutions follow from these equations in the
limit, when $\xi $ and $\eta $ tend to zero. The questions is
whether this limit depends on a relation between $\xi >0 $ and
$\eta >0$.

The viscous profile of stationary SW may locally be represented in
proper reference frame of the shock front by a stationary
continuous solution depending upon the only variable $x$; here
$\tau ^{\mu \nu} \to 0$ and all the parameters of this viscous
flow tend to constant values as $x\to\pm\infty$.

Without loss of generality we suppose further that the limiting
values of hydrodynamical parameters for $x \to -\infty $
correspond to the state ahead of the shock (denoted further by
index ``0'') and the values for $x \to +\infty $ correspond to the
state behind the shock (denoted further by index "1"), then we
have $u^{1}>0$
 behind and ahead of the shock.

Because all the values in (\ref{eq2}),(\ref{eq3}),(\ref{eq4}) depend only upon variable $x$, we have
from these equations

\begin{equation}
\label{eq5}
T^{1\nu}  + \tau ^{1\nu}  = T_{(0)}^{1\nu}  ,
\end{equation}

\begin{equation}
\label{eq6} u^{1}h^{\nu}  - h^{1}u^{\nu}  = H^{\nu}  \equiv
u_{(0)}^{1} h_{(0)}^{\nu}  - h_{(0)}^{1} u_{(0)}^{\nu},
\end{equation}

\begin{equation}
\label{eq7} nu^{1} = n_{(0)} u_{(0)}^{1}.
\end{equation}

As a result of $\tau ^{1\nu}  \to 0$ for $x \to \pm \infty $,
relations (\ref{eq5})-(\ref{eq7}) must be fulfilled for
corresponding asymptotic values $T^{\mu \nu} , n, h^{\mu} $,
obtained from continuous solutions of the system
(\ref{eq9})-(\ref{eq11}). Similarly to classical hydrodynamics
\cite{Rozhe} we interpret the conditions for shock transition from
the state $u_{(0)}^{\mu} ,h_{(0)}^{\mu}  ,n_{0} ,p_{0} $ (ahead of
the shock) into the state $u_{(1)}^{\mu} ,h_{(1)}^{\mu} ,n_{1}
,p_{1} $ (behind the shock), and as consequence of
(\ref{eq5})-(\ref{eq7}) these states must satisfy

\begin{equation}
\label{eq8} T_{(1)}^{1\nu}  = T_{(0)}^{1\nu}  ,
\end{equation}

\begin{equation}
\label{eq9} u_{(1)}^{1} h_{(1)}^{\nu}  - h_{(1)}^{1} u_{(1)}^{\nu}
= H^{\nu}  \equiv u_{(0)}^{1} h_{(0)}^{\nu} - h_{(0)}^{1}
u_{(0)}^{\nu}  ,
\end{equation}

\begin{equation}
\label{eq10} n_{(1)} u_{(1)}^{(1)} = n_{(0)} u_{(0)}^{1}  .
\end{equation}

\section{Dynamical system for the shock structure}

In this section we use some of the results of
\cite{ZhTytPLA,ZhTytZhTJE}. Suppose that equations
(\ref{eq8})-(\ref{eq10}) are fulfilled.

\textbf{Definition.} We say that shock transition $u_{(0)}^{\mu }
,h_{(0)}^{\mu}  ,n_0 ,p_0  \to u_{(1)}^{\mu} ,h_{(1)}^{\mu} ,n_1
,p_1 $ has viscous profile if there is a continuous solution of
(\ref{eq5})--(\ref{eq7}) having corresponding asymptotics for $x
\to -\infty $ and $x \to +\infty $.

We use the reference frame such that $u^{3}\equiv0$ and $h^{3}
\equiv 0$; $u^{1}$, $h^{1}$ being normal components to the surface
$x=const$, and $u^{2}$, $h^{2} $ being the tangential components;
 $u_{(0)}^{2} = 0$.

Following \cite{ZhTytPLA,ZhTytZhTJE}, due to (\ref{eq6}) we
represent $h^{\mu} $ in terms of $u^{1}$ and $u^{2}$:

\begin{equation}
\label{eq11} h^{\mu}  = {\frac{{1}}{{u^{1}}}}{[ {H^{\mu}  -
u^{\mu} H^{\alpha }u_{\alpha}}   ]}  .
\end{equation}

Multiplying (\ref{eq5}) by $u_{\nu} $ and taking into account that
$\tau ^{\mu \nu }u_{\nu} =0 $ we have

\begin{equation}
\label{eq12} \varepsilon ^{*} u^{1} = T_{(0)}^{1\mu}  u_{\mu}  ,
\end{equation}

\noindent
whence $\varepsilon $ and $h^{\mu} $ can be expressed through $u^{1}$ and
$u^{2}$.

Taking into account the explicit form of $\tau ^{\mu \nu} $ we
have from (\ref{eq5}) for $\nu =1,2$

\begin{equation}
\label{eq13}
 - \left( \xi + \frac{4}{3}\eta  \right)\left[1 +
\left(u^{1}\right)^{2}\right]\frac{\partial u^{1}}{\partial x} =
T_{(0)} ^{11} - T^{11},
\end{equation}

\begin{equation}
\label{eq14}
 - \eta \left[1 + \left(u^{1}\right)^{2}\right]\frac{\partial u^{2}}{\partial x} - \left(
\xi + \frac{\eta} {3} \right)u^{1}u^{2}\frac{\partial
u^{1}}{\partial x} = T_{(0)} ^{12} - T^{12}
\end{equation}

After eliminating of $\partial u^{1}$/$\partial x $ from
(\ref{eq14}) with the help of (\ref{eq13}) the second equation
transforms to

\[
 - \eta \left[1 + \left(u^{1}\right)^{2}\right]\frac{\partial u^{2}}{\partial x} = -
\frac{u^{1}u^{2}}{1 + \left(u^{1}\right)^{2}}\frac{\left( \xi +
\eta / 3 \right)}{\left( \xi + 4\eta / 3 \right)}\left( {T_{(0)}
^{11}} - {T^{11}} \right) + T_{(0)} ^{12} - T^{12}
\]

It is convenient to introduce a new variable $v = {{u^{2}}
\mathord{\left/ {\vphantom {{u^{2}} {\sqrt {1 +
\left(u^{1}\right)^{2}}}} } \right. \kern-\nulldelimiterspace}
{\sqrt {1 + \left(u^{1}\right)^{2}}}} $ in (\ref{eq13}) and
(\ref{eq14}); this yields dynamical system with respect to $u^{1}$
and $v$

\begin{equation}
\label{eq15}
\left( \xi + \frac{4}{3}\eta\right)\frac{du^{1}}{dx}
= F_{1} \left(u^{1},v\right),  \quad \quad    \eta \frac{dv}{dx} =
F_{2} \left(u^{1},v\right),
\end{equation}

\noindent
where

\[
F_{1} (u^{1},v) = p - {\frac{{1}}{{1 + (u^{1})^{2}}}}{\left[
{T_{(0)}^{11} + {\frac{{\mu}} {{4\pi}} }(H^{\alpha} u_{\alpha}
)^{2} - T_{(0)}^{1\mu} u_{\mu}  u^{1}} \right]} +
\]

\begin{equation}
\label{eq16}
+ \frac{\mu} {8\pi (u^{1})^{2}}[(H^{\alpha}
u_{\alpha} )^{2} - H^{\alpha} H_{\alpha}]
\end{equation}

\[
F_{2} (u^{1},v) = {\frac{{(T_{(0)}^{10} u^{0} - T_{(0)}^{12}
u^{2})u^{2}u^{1} - (\mu / 4\pi )u^{2}(H^{\alpha} u_{\alpha}
)^{2}}}{{u^{1}[1 + (u^{1})^{2}]^{{\frac{{5}}{{2}}}}}}} +
\]

\begin{equation}
\label{eq17}
 + {\frac{{(\mu / 4\pi )H^{2}(H^{\alpha} u_{\alpha}  ) - T_{(0)}^{12}
u^{1}}}{{[1 + (u^{1})^{2}]^{3 / 2}u^{1}}}}
\end{equation}

Continuous solutions of (\ref{eq15}) describe the SW structure. It
is important to note that $p(\varepsilon ,n)$ disappears from
$F_{2}$.

\section{Existence conditions of SW viscous profile.}

Let the state parameters $u_{(0)}^{\mu}, h_{(0)}^{\mu} ,n_{0}
,p_{0} $ ahead of the shock and $u_{(1)}^{\mu} ,h_{(1)}^{\mu}
,n_{1} ,p_{1} $ behind the shock satisfy the conservation laws
(\ref{eq8})--(\ref{eq10}) that relate hydrodynamic quantities on
both sides of SW. We denote $y=u^{1}$, $y_{0}=u_{(0)}^{1} $,
$y_{1}=u_{(\ref{eq1})}^{1} $.
% In this section we deal with the state variables $u^{\mu} $,$h^{\mu} $,$^{} n$,$ p$  ahead of the
% shock unless otherwise stated, and we omit further the index "0"
%  for these variables.

\begin{figure}
\includegraphics[height=0.5\textheight]{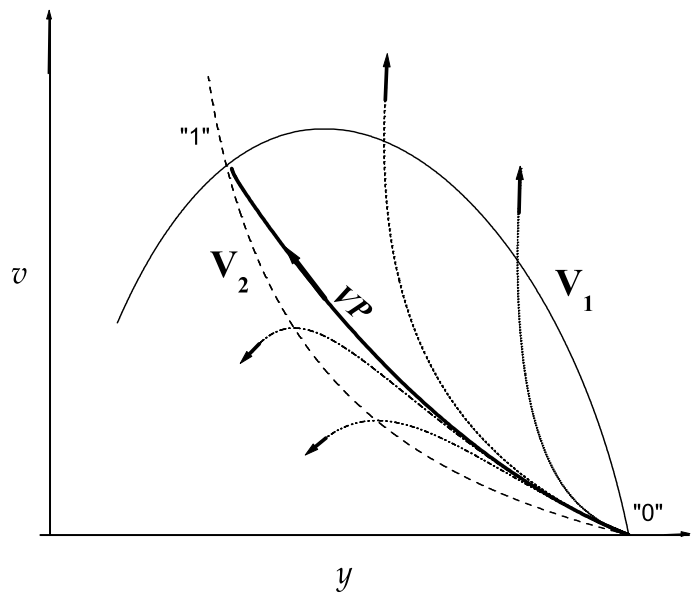}
\caption{The phase trajectories of  system (\ref{eq15})  in case
of DF; EOS is $p=\varepsilon/3$. The solid line ($\bf{V_1}$)
corresponds to $v=V_1(y)$, the dashed line ($\bf{V_2}$)
corresponds to $y=Y_2(v)$; the curve $VP$ connecting "0" and "1"
describes the viscous profile of the fast shock transition $"0"
\to "1"$. The arrows show the direction of the phase flow.}
\label{fast}
\end{figure}

\begin{figure}
\includegraphics[height=0.5\textheight]{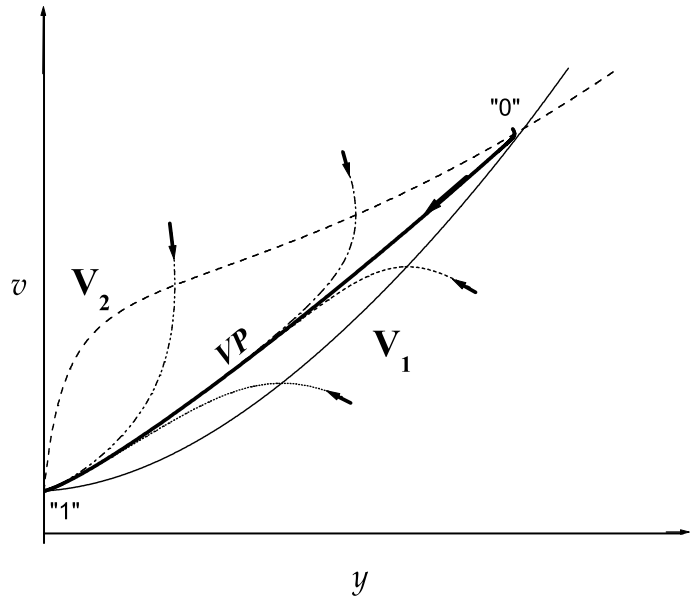}
\caption{The phase trajectories of  system \ref{eq15} in case DS
with the same EOS. In this case the curve $\bf{V_1}$ (solid) is
below $\bf{V_2}$ (dashed). The separatrix $VP$ of the saddle point
"0" describing the the viscous SW profile of slow shock transition
$"0" \to "1"$ goes to the final state "1". }
 \label{slow}
\end{figure}

Consider now the curves on $(y,v)$ -- plane where the right-hand
sides (\ref{eq16}), (\ref{eq17}) of system
 (\ref{eq15}) may change their signs.
 We denote $\bf{ V_1}$ a locus of points $(y,v)$ such that $F_{1} (y,v) =
 0$ and
$\bf{V_2}$ is a locus of points $(y,v)$ such that $F_{2} (y,v) =
 0$. We suppose that $\bf{V_1}$ is represented by a smooth
 function
 $v=V_{1}(y)$. From the results of \cite{ZhTytPLA,ZhTytZhTJE} it
 follows that  $\bf{V_2}$ is represented by a single-valued function
 $y=Y_{2}(v)$; this function may be not monotonous so it is more
 convenient to use this function instead of  the the inverse one.
 We suppose that "0" and "1" are connected by smooth components of
 $\bf{V_1}$ and $\bf{V_2}$ (see, e.g., Fig.\ref{fast}-- Fig.\ref{slow} ).

We consider a part of $(y,v)$ -- plane between $\bf{V_1}$ and
$\bf{V_2}$ such that there are two intersection points "0" and "1"
corresponding to the states ahead and behind the shock:
$v_{1}=V_{1}(y_{1})$, $y_{1}=Y_{2}(v_1)$; $v_{0}=V_1(y_0)$,
$y_0=Y_{2}(v_0)$, but the curves do not intersect between "0" and
"1".

The points $(y_0,V_1(y_0))$, $(y_1,V_1(y_1))$ are the rest points
of  system (\ref{eq15}).

We shall consider the following conditions.

A. The function $v=V_1(y)$ is a single-valued function on
$(y_1$,$y_0)$.

B. For all points of $\bf{V_2}$, $v \in (v_1,v_0)$ the following
inequality is valid (cf. \cite{ZhTytPLA,ZhTytZhTJE}):

\begin{equation}
\label{eq18} (y_{0} - y_{1} )F_{1} (Y_2(v),v) < 0  .
\end{equation}

Here we do not consider occurence of the Chapmen-Joguet points,
where the left hand side of (\ref{eq18}) equals to zero but does
not change its sign. This case may be studied by taking a
corresponding limit in (\ref{eq18}).

C. We suppose that $h^{1}h^{2} \ne $0 at the point "0".

This is a technical requirement. Otherwise we deal with much more simple
situation of parallel or perpendicular MHD SW; this case is not considered
here.

Let $u_{sl}$, $u_{f}$, $u_{A} $ stand for the speeds of
relativistic slow, fast and Alfven waves \cite{Lich,Sibg};
$u_{sl}$, $u_{f} $ being the roots of the polynomial $Q(y)$, where

\[
Q(y) = (1 - c_{S}^{2} )(y^{2} - u_{f}^{2} )(y^{2} - u_{sl}^{2} ) =
\]
\begin{equation}
\label{eq19} =  (1 - c_{S}^{2} )y^{4} - y^{2}\left( {c_{S}^{2} +
\left. {{\frac{{\mu {\left| {h} \right|}^{2}}}{{4\pi \left( {p +
\left. {\varepsilon}  \right)} \right.}}}} \right)} \right. +
{\frac{{\mu c_{S}^{2} (h^{1})^{2}}}{{4\pi \left( {p + \left.
{\varepsilon}  \right)} \right.}}}
\end{equation}

 $c_{S}^{2} = $ ($\partial $p/$\partial \varepsilon )_{S}$ is the speed of
sound.

The relativistic Alfven speed $u_{A}$ is defined by the formula

\begin{equation}
\label{eq20}
u_{A}^{2} = {\frac{{\mu (h^{1})^{2}}}{{4\pi (p^{\ast}  + \varepsilon ^{\ast
})}}}
\end{equation}

In \cite{ZhTytPLA,ZhTytZhTJE} one more parameter $u_{A}^{\ast}  $
has been introduced, which is the positive root of the equation
$R^{\ast} (y)=0$, where

\begin{equation}
\label{eq21}
R^{\ast} (y) = \left( {p + \left. {\varepsilon}  \right)} \right.y^{2}\left(
{1 + \left. {y^{2}} \right)} \right. - {\frac{{\mu}} {{4\pi}} }\left(
{h^{1}} \right)^{2}
\end{equation}

The above velocities satisfy the inequalities \cite{ZhTytZhTJE}

\begin{equation}
\label{eq22} u_{sl} < u_{A} < u_{A}^{\ast}  < u_{f} .
\end{equation}

We introduce one more requirement (one can show that it is
consistent with A,B).

D. Either

(DF): $u^{1}>u_{f}$ ahead of the shock at the point "0",

or

(DS): $u_{A}>u^{1}>u_{sl}$ ahead of the shock at the point "0" .

These inequalities correspond to the evolutionarity criteria of
classical MHD for the velocities ahead of the shock \cite{LLEl}.
The first inequality (DF) corresponds to the fast SW and the
second one (DS) -- to the slow SW.

By means of (\ref{eq10}) -- (\ref{eq12}) the variables $p$ and
$\varepsilon $ can be expressed in terms of the velocity
components $u^{1}$ and $u^{2}$.

\textbf{Lemma 1}. If  relations (\ref{eq7}), (\ref{eq11}) --
(\ref{eq12}) are satisfied, then at the point "0" we have

\[
{\left. {{\frac{{\partial S}}{{\partial u^{1}}}}} \right|}_{0} = {\left.
{{\frac{{\partial S}}{{\partial u^{2}}}}} \right|}_{0} = 0,
\]

\noindent
where $S$ is the entropy per baryon.

The proof uses the thermodynamical relation $TdS = pd(1 / n) + d(\varepsilon
/ n)$, where $T$ is the temperature. The statement of the lemma is obtained
after direct calculation of differentials \textit{dn} and \textit{dp} using (\ref{eq10})--(\ref{eq12}).

Using Lemma 1, after some calculations we have at the the point
"0" ($v_{0}=0$):

\begin{equation}
\label{A}
{\left. {{\frac{{\partial F_{1}}} {{\partial u^{1}}}}}
\right|}_{0} = {\frac{{(p + \varepsilon
)}}{{u^{1}(u^{0})^{4}}}}D(u^{1}),
\end{equation}

\noindent
where

\[
D(y) = (1 - \left. {c_{S}^{2}}  \right)y^{4} - y^{2}\left( {2c_{S}^{2} +
{\frac{{\mu {\left| {h} \right|}^{2}}}{{4\pi \left( {p + \left. {\varepsilon
} \right)} \right.}}} - 1} \right) + {\frac{{\mu [(h^{0})^{2} -
(h^{2})^{2}]}}{{4\pi \left( {p + \left. {\varepsilon}  \right)} \right.}}} -
c_{S}^{2} .
\]

The other derivatives of the right hand sides (\ref{eq16}),
(\ref{eq17}) of the dynamical system (\ref{eq15})  at the point
"0" are

\begin{equation}
\label{B}
 {\left. {{\frac{{\partial F_{1}}} {{\partial v}}}}
\right|}_{0} = {\frac{{\mu}} {{4\pi}}
}{\frac{{h^{1}h^{2}}}{{u^{0}u^{1}}}},
\end{equation}

\begin{equation}
\label{C}
 {\left. {{\frac{{\partial F_{2}}} {{\partial u^{1}}}}}
\right|}_{0} = {\frac{{\mu}} {{4\pi}}
}{\frac{{h^{1}h^{2}}}{{(u^{0})^{5}u^{1}}}},
\end{equation}

\begin{equation}
\label{D}
 {\left. {{\frac{{\partial F_{2}}} {{\partial v}}}}
\right|}_{0} = {\frac{{1}}{{(u^{0})^{4}u^{1}}}}R^{\ast} (u^{1}).
\end{equation}

Taking into account these relations in the vicinity of the point
"0" we have on the curve $\bf{V_2}$

\begin{equation}
\label{eq23}
F_{1} (y,V_{2} (y)) = {\frac{{\left( {p + \left. {\varepsilon}  \right)^{2}}
\right.}}{{u^{1}}}}{\frac{{{\rm Q}{\rm (}u^{1})}}{{R^{\ast} {\rm
(}u^{1})}}}(y - y_{0} ).
\end{equation}

On the curve $V_{2}$ we have

\begin{equation}
\label{eq24}
{\left. {{\frac{{dv}}{{dy}}}} \right|}_{0} = - {\frac{{\mu
h^{1}h^{2}}}{{4\pi u^{0}R^{\ast} (u^{1})}}}     ,
\end{equation}

\textbf{Lemma 2}. In the case DF the rest point "1" of system
(\ref{eq15}) is a saddle point. In the case DS the rest point "0"
is a saddle point.

Proof. Direct calculation yields

\[
{\frac{{\partial F_{1}}} {{\partial u^{1}}}}{\frac{{\partial F_{2}
}}{{\partial v}}} - {\frac{{\partial F_{1}}} {{\partial
v}}}{\frac{{\partial F_{2}}} {{\partial u^{1}}}} = {\frac{{(p +
\varepsilon )^{2}}}{{\left( {u^{1}} \right)^{2}\left( {u^{0}}
\right)^{4}}}}{ Q}{\rm (}u^{1}{\rm )}
\]

In the case (DS) ahead of SW we have $Q<0$ at "0" (see
(\ref{eq19})) and this yields the required statement according to
the properties of the saddle point \cite{BauLeo} . In the case
(DF) analogous result at "1" can be checked directly; however it
is easier to use the same relations as (\ref{A})-(\ref{D}) in case
of "0" by using the Lorents transformation that preserves
$u_{(0)}^{1}$ and transforms the transversal velocity component
$u_{(1)}^{2}$ to zero at the point "1".

Further the coefficients of $Q(y)$ and $R^{\ast} (y)$ involving the magnetic
field, energy density and baryon density are taken only at the point "0"
ahead of the shock.

Now we proceed to prove the existence of viscous profile.

Consider first the case (DF) for $y_{1}<y_{0}$, and put for
definiteness $h^{1}h^{2}>0$ at "0". Consideration of the opposite
sign of $h^{1}h^{2}$ is completely analogous.

First of all we note that the condition (\ref{eq18}) guarantees
that the trajectories of   system (\ref{eq15}) in the plane
$(y=u^1,v)$ cross the curve $\bf{V_2}$ from right to left (see
Fig. \ref{fast}).

On the curve $v=V_{1}(y)$:

\begin{equation}
\label{eq25}
F_{2} (y,V_{1} (y)) = - {\frac{{4\pi \left( {p + \left. {\varepsilon}
\right)} \right.^{2}{\rm Q}{\rm (}u^{1})}}{{\mu
h^{1}h^{2}u^{1}(u^{0})^{3}}}}(y - y_{0} ) > 0
\end{equation}

\noindent in the vicinity of the point "0", $y<y_{0}$.

Correspondingly trajectories of  system (\ref{eq15}) cross the
curve $V_{1}$ bottom-up (Fig. \ref{fast}). Evidently this is true
not only for the neighborhood of "0", but for all interval
($y_{1}$,$y_{0}$); otherwise there must be additional rest points
of  system (\ref{eq15}) between "0" and "1" , which contradicts to
our suppositions.

Now we shall find out relative disposition of $V_{1}$ and $V_{2}$.
Let at point "0"

\[
tg(\alpha _{1} ) = {\left. {{\frac{{dV_1}} {dy}}} \right|}_{0}
,\quad \quad {\rm }  tg(\alpha _2 ) = {\left. {{\frac{dV_2} {dy}}}
\right|}_{{\rm 0}}    .
\]

Using (\ref{A})-(\ref{B}) we obtain that the ratio of tangents at
"0" equals

\begin{equation}
\label{eq26}
{\frac{{tg(\alpha _{1} )}}{{tg(\alpha _{2} )}}} = {\frac{{16\pi ^{2}(p +
\varepsilon )^{2}(u^{0})^{2}}}{{(\mu h^{1}h^{2})^{2}}}}Q(u^{1}) + 1 =
{\frac{{16\pi ^{2}\left( {p + \left. {\varepsilon}  \right)} \right.}}{{(\mu
h^{1}h^{2}u^{0})^{2}}}}R^{\ast} (u^{1})D(u^{1}) \quad .
\end{equation}

Then in case of fast SW ($Q(y_{0})>0$) this ratio is $>1$, that is
$\bf{V_1}$ is above $\bf{V_2}$.

Therefore, the phase curves only can leave the domain between
$\bf{V_1}$ and $\bf{V_2}$ (Fig. \ref{fast}). Taking into account
the sign of $F_1$, it is easy to see that inside this domain all
the phase curves come out from the point "0" ; and there exists a
phase curve of (\ref{eq15}), that comes from "0" to "1" . Because
"1" is a saddle point (Lemma 2), this phase curve is unique. This
conclusion does not depend upon relation between (positive)
viscosity coefficients $\xi $, $\eta $.

Now we proceed to the case (DS) (see Fig. \ref{slow}); it is now
convenient to put $h^{1}h^{2}<0$ at "0" (the opposite case is
completely analogous). The sign of (\ref{eq25}) remains the same
as in the case (DF) and so is the direction of the phase curves
crossing $\bf{V_1}$. The direction of the phase curves crossing
$\bf{V_2}$ also remains due to (\ref{eq18}). Taking into account
(DS) and (\ref{eq22}) we have $Q(y_{0})<0$ and according to
(\ref{eq26}) we see that $\bf{V_2}$ is above $\bf{V_1}$ in the
neighborhood of "0".  As distinct from the case (DF) here the
phase curves can only enter the domain between "1" and "0".
Similarly to previous consideration there is a unique phase curve
of  system (\ref{eq15}), passing from "0" to "1"; this is a
separatrix of saddle point "0".

In the previous consideration we supposed that $y_{1}<y_{0}$; this
corresponds to usual compression SW. In case of anomalous EOS the
rarefaction shocks are also possible
\cite{Rozhe,BGKZh,ZhTytPLA,ZhTytZhTJE}. The condition (\ref{eq18})
is applicable in this case as well, and the consideration is
completely similar.

Therefore we proved that equations (\ref{eq5})--(\ref{eq10}) have
a unique continuous solution that connects the states "0" and "1".

\textbf{Theorem}. Let the states $u_{(0)}^{\mu}$, $h_{(0)}^{\mu}$,
$n_{0}$, $p_{0}$ ahead of the shock and $u_{(1)}^{\mu}$,
$h_{(1)}^{\mu}$, $n_{1}$, $p_{1}$ behind the shock satisfy the
conservation equations (\ref{eq8})--(\ref{eq10}). If the
conditions (A--D) are satisfied, then the MHD shock transition
"0"$ \to $"1" has a unique viscous profile satisfying equations
(\ref{eq5})--(\ref{eq10}).

\begin{figure}[htbp]
\includegraphics[height=0.5\textheight]{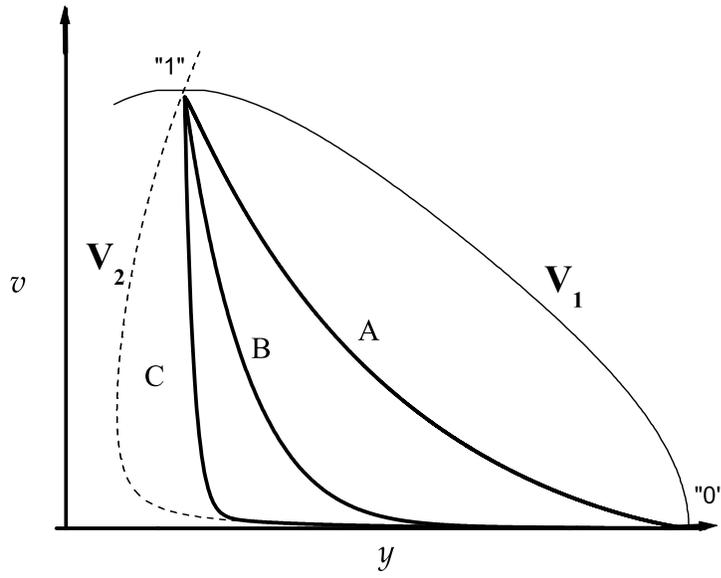}
\caption{The example of viscous SW profiles for three ratios of
$\eta/\xi$ in case of DF (fast SW) with non-monotonous dependence
$y=Y_2(v)$; $p=\varepsilon/3$. Disposition of curves $\bf{V_1}$
and $\bf{V_2}$ is as on Fig. \ref{fast}. The curve "A" describes
the profile with ratio $\eta/\xi=1$, "B" corresponds to
$\eta/\xi=0.1$ and "C" - to $\eta/\xi=0.01$. These curves
corresponding to the smaller ratios snuggle down to $\bf{V_2}$
(after they go out from "0") on some segment and then jump to "1".
} \label{eta-lim}
\end{figure}

Note that the analogous criteria for existence of SW viscous
profile  obtained in \cite{ZhTytPLA,ZhTytZhTJE} appear to be too
restrictive. These criteria have been obtained under condition
that one of the viscosity coefficients equals to zero ($\eta =0$),
and they rule out the shocks that satisfy the condition
$u_A<u^1<u^{*}_A$  at "1" after the shock front. However these
latter solutions are compatible with the criteria (A)-(D) of the
present paper. The explanation of this inconsistency is as
follows. If the function $Y_2(v)$ is monotonous and $\eta \to 0$,
then the phase curve of  system (\ref{eq15}) that go from "0" to
"1" tends to the curve $\bf{V_2}$. However, it may happen that
$Y_2(v)$ is not monotonous; this just corresponds to
$u_A<u^1<u^{*}_A$ at "1". In this case the above phase curve that
begins at "0" snuggles down to $\bf{V_2}$ only on some segment,
and the corresponding limiting solution for $\eta \to 0$ has a
discontinuity (see Fig. \ref{eta-lim}). This explains why such
solutions have been rejected in \cite{ZhTytPLA,ZhTytZhTJE},
because the initial supposition of these papers was the existence
of a regular viscous profile for $\eta =0$ .

\section{ Discussion}

The SW existence conditions in relativistic MHD with general
equation of state has been analyzed in \cite{ZhTytPLA,ZhTytZhTJE}
in case of $\eta =0$ in the Landau-Liftschits relativistic
viscosity tensor. In the present paper we obtained the criteria
for existence of viscous SW  profile dealing with both nonzero
viscosity coefficients ($\xi >0$, $\eta >0$) in this viscosity
tensor. If additional limitations on EOS (e.g., convexity) are
absent, our criteria are more restrictive than, e.g.,
evolutionarity conditions \cite{LLEl} or any other conditions that
involve characteristics of the fluid only at the initial and final
states. This is evident because condition (\ref{eq18}) must be
valid for the whole interval between the states "0" and "1".
 This situation is analogous to ordinary (non-magnetic)
hydrodynamics \cite{BGZh,BGKZh}; in this case our criteria reduce
to the criteria of these papers.

On the other hand, the criteria of the present paper are less
restrictive than that of \cite{ZhTytPLA,ZhTytZhTJE} obtained in
case of $\eta =0 $. This is because the supposition of existence
of a regular viscous profile used in \cite{ZhTytPLA,ZhTytZhTJE}
does not always hold in case of $\eta =0 $ (even if $\xi \ne 0$)
and this rules out some physical solutions. It should be noted
that such situation is specific just to relativistic MHD; this
does not appear neither in non-relativistic case, nor in ordinary
relativistic hydrodynamics.

Our criteria may be applied to arbitrary smooth EOS. However, we
must note that the requirement for $V_{1}(y)$ to be a continuous
(single-valued) function is not trivial and may not be fulfilled
in case of certain equations of state (cf., e.g., \cite{Rozhe}).
Though consideration of a viscous profile seems to be rather
effective for investigation of SW existence and stability, this
method may not work in case of complicated EOS (cf. remarks in
\cite{ZhTytpla}) that require either modification of equations of
the fluid motion or using additional physical information about
solutions.

\end{document}